\documentclass[fleqn,12pt]{article}
\usepackage{epsfig}
\begin{document}
\textheight 22cm
\textwidth 15cm
\noindent
{\Large \bf Predicting PDF tails of flux in plasma sheath region}
\newline
\newline
Johan Anderson\footnote{anderson.johan@gmail.com} and Eun-jin Kim$^{\star}$
\newline
Max-Planck-Institut f\"{u}r Plasmaphysik, IPP-Euratom Association, Teilinstitut Greifswald, 17491 Greifswald, Germany
\newline
$\star$ University of Sheffield, Department of Applied Mathematics, Hicks Building, Hounsfield Road, Sheffield, S3 7RH, UK
\newline
\begin{abstract}
\noindent
This letter provides the first prediction of the probability density function (PDF) of flux $R$ in plasma sheath region in magnetic fusion devices which is characterized by dynamical equations with exponential non-linearities. By using a non-perturbative statistical theory (instantons), the PDF tails of first moment are shown to be a modified Gumbel distribution which represents a frequency distribution of the extreme values of the ensemble. The non-Gaussian PDF tails that may be enhanced over Gaussian predictions are the result of intermittency caused by short lived coherent structures (instantons).
\end{abstract}
\newpage
\renewcommand{\thesection}{\Roman{section}}
In the edge region of magnetically confined fusion experiment the plasma exhibits large bursty fluctuations, often called intermittency~\cite{a11}-~\cite{a19}. Intermittent turbulence is characterized by patchy spatial structure that is bursty in time. The likelihood of these events are described by probability density functions (PDFs) which significantly deviate from Gaussian predictions. More specifically, exponential scalings are often observed in the PDF tails in magnetic confinement experiments~\cite{a16}-~\cite{a19}, and intermittency at the edge strongly influences the overall global particle and heat transport. In particular it may for instance influence the threshold for the high confinement mode (H-mode) in tokamak experiments~\cite{a20}. In the view of these experimental results, theories built on average transport coefficients and Gaussian statistics fall short in predicting vital transport processes. The high possibility of confinement degradation by intermittency strongly calls for a powerful predictive theory.

The purpose of this letter is to provide the theoretical prediction for the PDF tails of flux in plasma sheath by using a non-perturbative statistical method. The basic idea is to associate the bursty event with the creation of a coherent structure (e.g.  blobs, streamers etc.). The candidate that could describe the creation process is the instanton which is localized in time, existing during the formation of the coherent structure. This idea is embedded in our theoretical method - the so-called instanton method - which is a non-perturbative way of calculating the PDF tails~\cite{a21}-~\cite{a31}. Note that, the instanton method has been adopted from quantum field theory and then modified to classical statistical physics for Burgers turbulence~\cite{a22}-~\cite{a23} and in the model by Kraichnan~\cite{a32}.

Our previous papers using this instanton method have shown that the PDF tails of momentum flux and heat flux are significantly enhanced over the Gaussian prediction~\cite{a24}-~\cite{a28}, providing a novel explanation for exponential PDF tails of momentum flux found in recent experiments at CSDX at UCSD~\cite{a19}. PDF tails of forced shear flow $R$ in a non-linear diffusion model were shown to be $\sim \exp\{- c (R/R_0)^{4}\}$ that falls off much faster than a Gaussian by both analytical (e.g. instanton) and numerical investigations~\cite{a30}.

In the following we study the PDFs of flux of 1st moment variables such as potential, density etc in plasma sheath where electric potential is governed by exponential non-linear interaction. In Ref.~\cite{a291}, we computed the PDF tails of blob density by using a simplified model for blobby transport with different non-linear interaction. Note that in the instanton analysis a careful treatment of the dominant non-linear interaction is necessary so that keeping only leading order non-linear interaction by Taylor expansion may lead to the erroneous predictions for the PDF tails~\cite{a31}. By using the instanton method we show that the PDF tails of the first moment satisfy  modified Gumbel distributions~\cite{a33}-~\cite{a34}. The Gumbel distribution is used to model extreme value distributions, i.e. a frequency distribution of the largest or smallest values of the ensemble. This is due to the fact that Taylor expansion of the exponential contains infinite sum of all powers. In a comparison with a Gaussian distribution it is shown that the flux in plasma sheath may exhibit stretched PDF tails. We emphasise that the predicted Gumbel distribution is new and has not been found in any related previous work.

 As a physical model of plasma sheath, we consider a generic simplified model of turbulence with a dominant exponential non-linear interaction. The model is a simplified version of the three field Braghinskii fluid model presented in Ref.~\cite{a36} where we have retained the exponential non-linear interaction. Note that this model can be reduced to the model derived in Ref.~\cite{a35} by linearization and keeping only non-linear terms up to second order. This model is sufficient for the prediction of the PDF tails since our previous papers showed that the PDF tails are rather insensitive to the details of the dynamics and dependent only on the dominant non-linear interaction term~\cite{a31}. We model the sheath region with a general non-linear equation describing the time evolution of the variable $\phi$ for fluctuations with exponential interaction $K e^{c \phi}$~\cite{a17},~\cite{a35}-~\cite{a37},
\begin{eqnarray}
\frac{\partial \phi}{\partial t} + K e^{c \phi} = f.
\end{eqnarray}
Here $f$ is the forcing. The statistics of the forcing is assumed to be Gaussian with a short correlation time modeled by the delta function as
\begin{eqnarray}
\langle f(x, t) f(x^{\prime}, t^{\prime}) \rangle = \delta(t-t^{\prime})\kappa(x-x^{\prime}),
\end{eqnarray}
and $\langle f \rangle = 0$. The angular brackets denote the average over the statistics of the forcing $f$.
The delta correlation in time is chosen for the sake of simplicity of the analysis. Note that in the case of a finite correlation time non-local integral equations in time are needed.

We calculate the PDF tails of local flux $M(\phi(x=x_0))$ that is the first moment of $\phi$ (e.g. density, potential) by using the instanton method. The PDF tails are expressed in terms of a path-integral using the Gaussian statistics of the forcing~\cite{a21}. The optimum path is then associated with the creation of a short lived coherent structure (among all possible paths) and the action is evaluated using the saddle-point method on the effective action. The saddle-point solution of the dynamical variable $\phi(x,t)$ of the form $\phi(x,t) = F(t) v(x)$ is called an instanton if $F(t) = 0$ at $t=-\infty$ and $F(t) \neq 0$ at $t=0$.  Note that, the function $v(x)$ here represents the spatial form of the coherent structure. The probability density function of the flux $R$ can be defined as
\begin{eqnarray}
P(R) =  \langle \delta(M(\phi(x=x_0)) - R) \rangle = \int d \lambda e^{i \lambda R} I_{\lambda},
\end{eqnarray}
where 
\begin{eqnarray}
I_{\lambda} = \langle \exp(-i \lambda M(\phi(x=x_0))) \rangle.
\end{eqnarray}
The integrand can then be rewritten in the form of a path-integral as
\begin{eqnarray}
I_{\lambda} = \int \mathcal{D} \phi \mathcal{D} \bar{\phi} e^{-S_{\lambda}}.
\end{eqnarray}

Here the effective action $S_{\lambda}$ in Eq. (5) is expressed as,
\begin{eqnarray}
S_{\lambda} & = & -i \int dx dt \bar{\phi} \left( \frac{\partial \phi}{\partial t} + K e^{c \phi} \right) \nonumber \\
& + & \frac{1}{2} \int dx dx^{\prime} dt \bar{\phi}(x,t) \kappa(x-x^{\prime}) \bar{\phi}(x^{\prime},t) \nonumber \\
& + & i \lambda \int dx dt M(\phi)\delta(x-x_0) \delta(t).
\end{eqnarray}
We make use of a logarithmic transform $ \phi = \ln u$ and $u = v(F+1)$ where $v$ is the spatial structure of the instanton and $F$ is the time dependent function. The action can now be recast into,
\begin{eqnarray}
S_{\lambda} & = & -i \int dt c_1 \bar{F}_1 \left( \frac{\partial \ln(F+1)}{\partial t} + c_2(F+1)^c \right) \nonumber \\
& + & \frac{1}{2} c_1 c_4 \int dt \bar{F}_1^2 \nonumber \\
& + & i \lambda \int dt c_1 c_3 \ln(F+1) \delta(t).
\end{eqnarray}
Here we have used the definitions,
\begin{eqnarray}
c_1 & = & \int dx \bar{v}(x), \\
c_1 c_2 & = & \int dx \bar{v}(x) K v(x)^c, \\
c_1 c_3 & = & \int dx \delta(x-x_0), \\
c_1 c_4 & = & \int dx dy \bar{v}(x) \bar{v}(y) K(x-y),
\end{eqnarray}

In order to find the path with highest probability identified by the instanton or the extremum of the action we compute the first variational derivatives to minimize $S_{\lambda}$ with respect to $F$ and $\bar{F}_1$, 
\begin{eqnarray}
\frac{\delta S_{\lambda}}{ \delta \bar{F}_1} & = & -i\left(\frac{\partial}{\partial t} \ln(F+1) + c_2 (F+1)^c \right) + c_4 \bar{F}_1 = 0, \\
\frac{\delta S_{\lambda}}{ \delta F} & = & -i \left(\frac{- \dot{\bar{F}}_1}{F+1} + c_2 c (F+1)^{c-1}\bar{F}_1 \right) + i \lambda c_3 \frac{1}{F+1} \delta(t) = 0.
\end{eqnarray}
Eq. (12) gives
\begin{eqnarray}
+i \left( \dot{F} + c_2 (F+1)^{c+1} \right) = c_4 (F+1) \bar{F}_1
\end{eqnarray}
Differentiating Eq. (14) w.r.t time gives,

\begin{eqnarray}
\left(\ddot{F} + c_2 (c+1)(F+1)^c\dot{F} -\frac{\dot{F}^2}{F+1} - c_2 (F+1)^c \dot{F} \right) \nonumber \\
 =  -i c_4 (F+1) \dot{\bar{F}}_1.
\end{eqnarray}

We solve Eq. (15) for $t<0$ and match the solution at $t=0$. Note that the instanton solution $F$ rapidly grows at $t=0$ with increasing $\lambda$ while it vanishes as $t \rightarrow -\infty$. By using Eqs. (13) to (15),
\begin{eqnarray}
- \frac{d^2}{dt^2} \ln(F+1) +  c c_2^2 (F+1)^{2c} & = & 0,
\end{eqnarray}
we find the solution,
\begin{eqnarray}
\frac{d}{dt} \ln(F+1) & = & c_2 (F+1)^c, \\
F(t) & = & -1 + \left(\frac{\alpha_1}{c_2^2} + \frac{\sqrt{\alpha_1}}{c_2^2} \tanh (c \sqrt{\alpha_1} t - \alpha_2/2)\right)^{1/2c}.
\end{eqnarray}
The constants $\alpha_1$ and $\alpha_2$ are to be determined using the boundary conditions $F(0) = F_0$ and $F(-\infty) = 0$. To find the value of $F$ at $t=0$ we integrate Eq. (13) over $(-\epsilon, \epsilon)$ and use Eq. (14) to obtain, 
\begin{eqnarray}
\bar{F}_1(0) \approx c_3 \lambda.
\end{eqnarray}
We then use Eq. (12) at $t=0$ to obtain
\begin{eqnarray}
F(0)^c = F_0^c \approx -i \lambda \frac{c_4 c_3}{2 c_2}.
\end{eqnarray}
Using Eq. (20) and the condition that $F(-\infty) = 0$, we find the values of $\alpha_1 = \frac{1}{4}(1 \pm \sqrt{1-4 c_2^2})^2$ and $\alpha_2 = -2 \tanh^{-1} (\frac{c_2^2(F_0 + 1) + \alpha_1}{\sqrt{\alpha_1}})$.

We now find an approximate value of the PDF in the limit of $\lambda \rightarrow \infty$ which is expressed by a path-integral, using the saddle point method. First we compute the $\lambda$-dependence of the action $S_{\lambda}$ and then proceed to determine the PDF tails. In the limit of $\lambda \rightarrow \infty$, $S_{\lambda}$ becomes

\begin{eqnarray} 
S_{\lambda} & = & -i \int dt c_1 \left( \bar{F}_1 (\frac{\partial}{\partial t} \ln(F+1) + c_2 (F+1)^c) \right) \nonumber \\
& + & \frac{1}{2} c_1 c_4 \int dt \bar{F}_1^2 \nonumber \\
& + & i \lambda \int dt c_1 c_3 \ln(F+1) \delta(t) \nonumber \\
& = & \frac{2 c_1^2 c_2 }{c_4 c} F^c(0) + i \lambda c_3 \ln(F(0)+1) \nonumber \\
& = & - i k_1 \lambda + i k_1 \lambda \ln ( -i k_2 \lambda),
\end{eqnarray}
where the coefficient $k_1$ and $k_2$ are defined as
\begin{eqnarray}
k_1 & = & \frac{c_1 c_3}{c}, \\
k_2 & = & \frac{c_3 c_4}{2 c_2}.
\end{eqnarray}
The tail of the PDF is then found by performing the $\lambda$-integral in Eq. (3) by the saddle point method in the limit $R \rightarrow \infty$. It is later shown that this corresponds to $\lambda \rightarrow \infty$,
\begin{eqnarray}
P(R) & \sim & \int d \lambda e^{i \lambda R + i k_1 \lambda - i k_1 \lambda \ln (- i k_2 \lambda)} \sim \frac{1}{\sqrt{f^{\prime \prime}(\lambda_0)}} e^{f(\lambda_0)}\\
& \approx & e^{-\frac{k_1}{k_2} e^{R/k_1} + R/(2 k_1)}.
\end{eqnarray}
Here, we have evaluated the $\lambda$-integral using the extreme point $f^{\prime}(\lambda_0) = 0$ and $ \lambda_0 = i \frac{1}{k_2} e^{R/k_1}$ of $f(\lambda) = - i \lambda R + i k_1 \lambda + i k_2 \lambda \ln \lambda$  by using the saddle-point method. Note that the first term ($e^{-\frac{k_1}{k_2} e^{R/k_1} }$) in Eq. (25) is the zeroth order term and ($ e^{R/(2 k_1)}$) is the correction term. The resulting PDF tail in Eq. (25) satisfies a Gumbel distribution~\cite{a33}-~\cite{a34}. The Gumbel distribution is one type of an extreme value distribution that is the limit distribution of maxima/minima of independent and identically distributed random variables. Thus the Gumbel distribution is one model of approximate maxima/minima of finite sequences of random variables. This is due to the fact that Taylor expansion of the exponential contains an infinite sum of all powers.

We now show explicitly that systems with exponential non-linearities may have enhanced PDF tails compared to Gaussian distributions.
In figure 1, a comparison of the PDF tails of flux with an exponential non-linearity found in Eq. 25 with $k_1 = 2.0$ and $k_2 = 1.0$ (black line), $k_1 = 1.0$ and $k_2 = 1.0$ (green line), $k_1 = 0.5$ and $k_2 = 1.0$ (blue line) and a Gaussian distribution (red line) are shown. We have normalized all the distributions as one-sided PDFs. The figure shows that the PDF tails with exponential non-linearity may be enhanced over a Gaussian distribution depending on the parameters $k_1$ and $k_2$. It is important to note that as $R \rightarrow \infty$ the Gaussian distribution takes larger values than the Gumbel distribution the probability of taking such a large value of $R$ may become extremely small ($\sim 10^{-10}$). However, if the exponential non-linear interaction is weak and can be expanded in a truncated Taylor series up to second order the plasma sheath region is governed by a power-law non-linearity, a slightly different PDF tail $\sim \exp\{- c (R/R_0)^{s}\}$ with $s=(n+1)/m$ can be expected~\cite{a31}. Here, $n=2$ and $m$ are the order of the highest dominant nonlinear interaction term and moments for which the PDFs are computed, respectively.

\begin{figure}
  \includegraphics[height=.3\textheight]{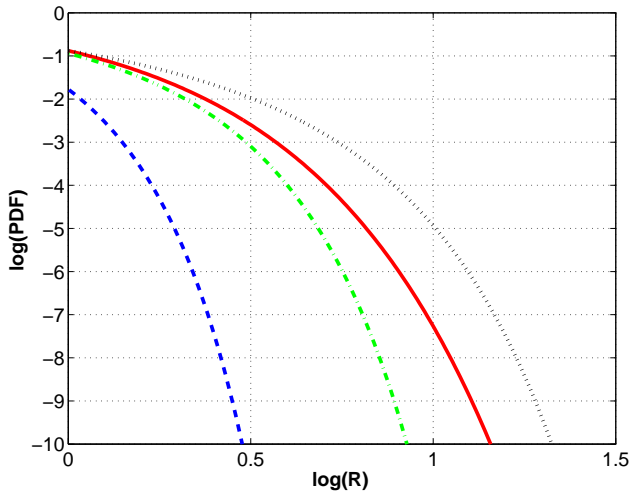}
  \caption{The PDF tails of flux in a system with exponential non-linearity with $k_1 = 2.0$ and $k_2 = 1.0$ (black line), $k_1 = 1.0$ and $k_2 = 1.0$ (green line), $k_1 = 0.5$ and $k_2 = 1.0$ (blue line) and a Gaussian distribution (red line) are shown.}
\end{figure}

Note that the usefulness of the found PDF tails is limited by the spatial structure of the specific solutions to the non-linear equations. The spatial structure determines the magnitude and sign of the coefficients $k_1$ and $k_2$.

Furthermore, note that to evaluate the accuracy of the Gumbel distribution in representing the PDF would require the exact computation of our path integral and $\lambda$-integral in Eq. (3), which is a rather formidable task beyond the scope of the present paper. We can however obtain rough estimate on errors in our predictions by considering the two major, leading order contributions that were neglected - the first one in using the saddle-point method for the path integral [Eq. (5)], and the second in the $\lambda$-integral in Eq. (24). We note that we computed the PDFs keeping zeroth, first and second order terms in the $\lambda$-integral and up to first order in using the saddle-point path integral. For instance, for $R=0.6$ an error of the same order as the expansion in inverse aspect ratio $\approx 30 \% $ (at $R=1$ the error is only a few percent) is found from the sub-leading order terms. Therefore, the Gumbel distribution is a reasonable estimate for $R>0.6-0.7$. The exact calculation of corrections to our model is however very involved, and will be addressed in future publications.

In summary, in the present letter we have presented a statistical theory of turbulence and intermittency due to coherent structures in the sheath region in fusion devices. The use of coherent structures is motivated by various experimental results in the sheath region that bursty events cause a significant transport which are linked to coherent structures. To investigate intermittency caused by these coherent structures, we have computed the PDF tails of 1st moment flux (e.g. density, potential) using the non-perturbative technique called the instanton method. The PDF tails are found to be in the form of modified Gumbel distributions. Interestingly, the Gumbel PDF tails may be enhanced in comparison with a Gaussian distribution. It is important to note that PDF tails of flux were shown to be far from Gaussian although Gaussian forcing was used. This result could guide in interpreting experimental results. Specifically, the PDF tails derived here and in previous papers may directly be compared with experimental results using log-log or log-lin plots. Furthermore this model could elucidate the very complex intermittent behavior at the edge of magnetic confinement systems. 
\vspace{0.5cm} \\

{\bf \Large Acknowledgment} \\
This research was supported by the Engineering and Physical Sciences Research Council (EPSRC) EP/D064317/1.


\newpage
\begin{thebibliography}{200}
\bibitem{a11} Stangeby P C 2000 The plasma boundary of magnetic fusion devices (IOP publishing, Bristol)
\bibitem{a12} Sattin F, Vianello N and Valisa M 2004 {\it Phys. Plasmas} {\bf11} 5032
\bibitem{a13} Carter T A 2006 {\it Phys. Plasmas} {\bf 13} 010701
\bibitem{a14} Saha S K and Chowdhury S 2006 {\it Phys. Plasmas} {\bf 13} 092512
\bibitem{a15} Agostini M, Zweben S J, Cavazzana R, Scarin P, Serianni G, Maqueda R J and Stotler D P 2007 {\it Phys. Plasmas} {\bf 14} 102305
\bibitem{a16} Zweben S J, Boedo J A, Grulke O, Hidalgo C, LaBombard B, Maqueda R J, Scarin P, and Terry J L 2007 {\it Plasma Phys. Controlled Fusion} 49 S1
\bibitem{a17} Sattin F, Scarin P, Agostini M, Cavazzana R, Serianni G, Spolaore M, and Vianello N 2006 {\it Plasma Phys. Controlled Fusion} 48 1033
\bibitem{a18} Myra J R, Russell D A, and DIppolito D A 2008 {\it Phys. Plasmas} 15 032304
\bibitem{a19} Yan Z, Tynan G R, Yu J H, Holland C, Muller S, and Xu M, 2007 {\it Bull. Am. Phys. Soc.} {\bf 52} 74
\bibitem{a20} Connor J W and Wilson H R 2000 {\it Plasma Phys. Contr. Fusion} {\bf 42} 1408 
\bibitem{a21} Zinn-Justin J 1989 Field Theory and Critical Phenomena (Clarendon, Oxford)
\bibitem{a22} Gurarie V and Migdal A 1996 {\it Phys. Rev. E} {\bf 54} 4908
\bibitem{a23} Falkovich G, Kolokolov I, Lebedev V and Migdal A 1996 {\it Phys. Rev. E} {\bf 54} 4896
\bibitem{a24} Kim E and Diamond P H 2002 {\it Phys. Plasmas} {\bf 9} 71
\bibitem{a25} Kim E and Diamond P H 2002 {\it Phys. Rev. Lett.} {\bf 88} 225002
\bibitem{a26} Kim E, Diamond P H, Malkov M, Hahm T S, Itoh K, Itoh S, Champeaux S, Gruzinov I, Gurcan O, Holland C, Rosenbluth M N and Smolyakov A 2003 {\it Nucl. Fusion} {\bf 43} 961
\bibitem{a27} Anderson J and Kim E 2008 {\it Phys. Plasmas} {\bf 15} 052306
\bibitem{a28} Anderson J and Kim E 2008 {\it Phys. Plasmas} {\bf 15} 082312
\bibitem{a29} Anderson J and Kim E 2009 {\it Nuclear Fusion} {\bf 49} 075027
\bibitem{a291} Anderson J and Kim E 2008 {\it Phys. Plasmas} {\bf 15} 122303 
\bibitem{a30} Kim E, Liu HL and Anderson J 2009 {\it Phys. Plasmas} {16}, 052304
\bibitem{a31} Kim E and Anderson J 2008 {\it Phys. Plasmas} {\bf 15} 114506
\bibitem{a32} Balkovsky E and Lebedev V 1998 {\it JETP Lett.} 68 616
\bibitem{a33} Fisher R A and Tippett L H C 1928 {\it Proc. Cambridge Philosophical Society} 24 190
\bibitem{a34} Gumbel E J 1958 Statistics of Extremes (Columbia University Press, New York)
\bibitem{a36} D'Ippolito D A, Myra J R and Krasheninnikov S I 2002 {\it Phys. Plasmas} {\bf 9} 222
\bibitem{a35} Krasheninnikov S I 2001 {\it Phys. Lett. A} {\bf 283} 368
\bibitem{a37} Sarazin Y and Ghendrih P 1998 {\it Phys. Plasmas} {\bf 5} 4214
\end{thebibliography}
\end{document}